\documentclass[twocolumn,showpacs,preprintnumbers,amsmath,amssymb]{revtex4}
\input psfig.sty

\usepackage{graphicx}
\usepackage{dcolumn}
\usepackage{bm}

\begin{document}

\def\be{\begin{equation}}
\def\ee{\end{equation}}


\title{Nonextensive models for earthquakes}

\author{R. Silva} \email{rsilva@uern.br,rsilva@on.br}
\affiliation{Observat\'orio Nacional, Rua Gal. Jos\'e Cristino 77,
20921-400 Rio de Janeiro - RJ,Brasil}
\affiliation{Universidade do Estado do Rio Grande do Norte,
59610-210, Mossor\'o, RN, Brasil}

\author{G. S. Fran\c{c}a} \email{george@dfte.ufrn.br}
\affiliation{Departamento de F\'\i sica, Universidade Federal do
R. G. do Norte, 59072-970, Natal, RN, Brasil}

\author{C. S. Vilar} \email{vilar@dfte.ufrn.br}
\affiliation{Departamento de F\'\i sica, Universidade Federal do
R. G. do Norte, 59072-970, Natal, RN, Brasil}

\author{J. S. Alcaniz} \email{alcaniz@on.br}

\affiliation{Observat\'orio Nacional, Rua Gal. Jos\'e Cristino 77,
20921-400 Rio de Janeiro - RJ,Brasil}

\date{\today}

\begin{abstract}
We have revisited the fragment-asperity interaction model recently introduced by Sotolongo-Costa and Posadas 
(Physical Review Letters 92, 048501, 2004) \cite{oscar2004} by considering a different definition for mean 
values in the context of Tsallis nonextensive statistics and introducing a new scale between the 
earthquake energy and the size of fragment $\epsilon \propto r^3$. The energy distribution function (EDF) deduced in our approach is 
considerably different from the one obtained in the above reference. We have also tested the viability of this 
new EDF with data from two different catalogs (in three different areas), namely, NEIC and Bulletin Seismic of 
the Revista Brasileira de Geof\'{\i}sica. Although both approaches provide very similar values for the nonextensive 
parameter $q$, other physical quantities, e.g., the energy density differs considerably, by several orders of magnitude.
\end{abstract}

\pacs{89.75.Da;91.30.Bi;91.30.Dk}
\maketitle

\section{Introduction}

Over the last two decades, a great deal of attention has been paid to the so-called nonextensive Tsallis entropy, 
both from theoretical and observational viewpoints. This particular nonextensive formulation \cite{T88,SL99} seems 
to present a consistent theoretical tool to investigate complex systems in their nonequilibrium stationary states, systems 
with multifractal and self-similar structures, systems dominated by long-range interactions, anamolous diffusion phenomena, 
among others. Some recent applications of Tsallis entropy $S_{q\neq 1}$ to a number of complex scenarios is now providing a 
more definite picture of the kind of physical problems to which this $q$-formalism can in fact be applied.

In this regard, systems of interest in geophysics has also been studied in light of this nonextensive formalism. In this 
particular context, the very first investigation was done by Abe \cite{abe03a} who showed that the statistical 
properties of three-dimensional distance between successive earthquakes  follow a $q$-exponential function with the 
nonextensive parameter lying in the interval $[0,1]$ \cite{abe01}. Since then, other geophysical analyses have been 
performed as, for instance, the statistics of the calm time, which indicates a scale-free nature for earthquake 
phenomena and corresponds to a $q$-exponential distribution with $q>1$ \cite{abe04}, and models for temperature 
distributions and radon emission of volcanos \cite{gerv04}. More recently, a very interesting model for earthquakes 
dynamics related to Tsallis nonentensive framework has been proposed by Sotolongo-Costa and Posadas 
(SCP Model) \cite{oscar2004}. Such a model consists basically of two rough profiles interacting via fragments filling 
the gap between them, where the fragments are produced by local breakage of the local plates.  By using the nonextensive 
formalism the authors of Ref. \cite{oscar2004} not only showed the influence of the size distribution of fragments on 
the energy distribution of earthquakes but also deduced a new energy distribution function (EDF), which gives 
the well-known Gutenberg-Richter law \cite{gut44} as a particular case.

However, in dealing with this nonextensive framework, a particular attention must be paid to the possible definitions 
for mean values, which play a fundamental role within the domain of this nonextensive statistics \cite{tsallis98}. 
In this concern, recent studies of the properties of the relative entropy and the Shore-Johnson theorem for consistent 
minimum cross-entropy principle, revealed the necessity of the so-called $q$-expectation value in studies involving 
this nonextensive statistical mechanics (see \cite{abe05} for details). Thus, by introducing this $q$-definition of 
mean value we re-analyzed the fragment-asperity interaction model of Sotolongo-Costa and Posadas \cite{oscar2004}. 
Moreover, a new scale law between the released relative energy $\epsilon$ and the $3$-dimensional size of 
fragments has also been introduced. By using the standard method of entropy maximization we also deduced a 
new energy distribution function, which differs considerably from the one obtained in Ref. \cite{oscar2004}. In order to 
test the viability of our appoach we used data taken from two seismic catalogs, namely, NEIC and Bulletin Seismic of the 
Revista Brasileira de Geofisica. It is shown that although both approaches provide very similar values for the nonextensive 
parameter $q$, the other physical quantity, e.g., the energy density differ by several orders of magnitude.
 
This paper is organized as follows. In Sec. II, the standard formalism of nonextensive statistical mechanics is reexamined, 
as well as the theorical basis of SCP model. In Sec III, a new EDF is analytically calculated through extremization 
of Tsallis' entropy under the constrains of the $q$-expectation value and normalization condition. In Sec IV, we test 
this new EDF with data from two different catalogs and estimate the best-fit values for the nonextensive parameter $q$ 
and the proportionality constant between the released relative energy $\epsilon$ and the volume of the fragments $r^3$, 
i.e., the energy density, $a$. We end this paper by emphasizing the main results in the 
conclusion Section.

\section{Non-extensive framework and SCP model}

In this Section, we recall the nonextensive theoretical basis of the SCP model. As widely known, the Tsallis' statistics generalizes the Botzmann-Gibbs statistics in what concerns the concept of entropy. Such formalism is based on the parametric class of entropies given by
\begin{equation} \label{eq:1}
S_{q\neq 1}=-k_B\int p^q(\sigma)\ln_q p(\sigma) d\sigma ,
\end{equation}
where  $k_{B}$ is the Boltzmann constant. In the SCP model, $p(\sigma)$ stands for the probability of finding a fragment of relative surface $\sigma$ (which is defined as a characteristic surface of the system), $q$ is the nonextensive parameter and the $q$-logarithmic function above is defined by
\begin{equation}\label{eq:224}
\ln_q p\, =\, (1-q)^{-1}(p^{1-q}-1), \,\,\,\,\,\,\,\,\, (p>0)
\end{equation}
which recovers the standard Boltzmann-Gibbs entropy $S_1 = - k_{B}\int p \ln p {d^{3}p}$ in the limit $q\rightarrow 1$.
It is worth mentioning that most of the experimental evidence supporting Tsallis proposal are related to the power-law distribution associated with $S_{q\neq 1}$ descripition of the classical $N$-body problem \cite{SPL98}.

The SCP model is a simple approach for earthquakes dynamics revealing a very interesting application of the Tsallis' framework. Indeed, the fundamental idea consists in the fact that the space between faults is filled with the residues of the breakage of the tectonic plates. In this regard, the authors studied the influence of the size distribution of fragments on the energy distribution of earthquakes. The theoretical motivation follows from the fragmentation phenomena \cite{englman87} in the context of the geophysics systems. In this latter work, Englaman et al showed that the standard Botzmann-Gibbs formalism, although useful, cannot account for an important feature of fragmentation process, i.e., the presence of scaling in the size distribution of fragments, which is one of the main ingredients of the SCP approach. Thus, a nonextensive formalism is not only justified in SCP model but also necessary since the process of violent fractioning is very probably a nonextensive phenomenon, leading to long-range interactions among the parts of the object being fragmented (see, e.g., \cite{oscar2004,oscar2000}). In reality, such an influence was earlier emphasized in other investigations \cite{sauler96}. In general lines, the SCP model follows similar arguments to those presented in Refs. \cite{bhv96} being, however, a more realistic seismic model than the one proposed in Ref. \cite{herrmann90}. In particular, the theoretical ingredients reads:
\begin{itemize}
 
\item the mechanism of relative displacement of fault plates is the main cause of earthquakes;
 
\item the surfaces of the tectonic plates are irregular and the fragments filling the space between them are very diverse and have irregular shapes;

\item the mechanism of triggering earthquakes is established through the combination of irregularities of the fault planes and the distribution of fragments between them;
 
\item the fragment distribution function and consequently the EDF emerges naturally from a nonextensive framework.

\end{itemize}
 
From the above arguments, the EDF deduced in Ref. \cite{oscar2004} is given by
\begin{eqnarray}
\log (N_{>m}) & = & \log N + \left(\frac{2-q}{1-q}\right) \times \\ \nonumber & & \times \log \left[1 + a(q-1)  (2-q)^{(1-q) \over (q-2)}\times 10^{2m}\right].  
\end{eqnarray}
According to Ref. \cite{oscar2004}, the above expression describes very well the energy distribution in all detectable range of magnitudes, unlike the empirical formula of Gutenberg-Richter \cite{gut44}.

\begin{figure*}[t]
\vspace{.2in}
\centerline{\psfig{figure=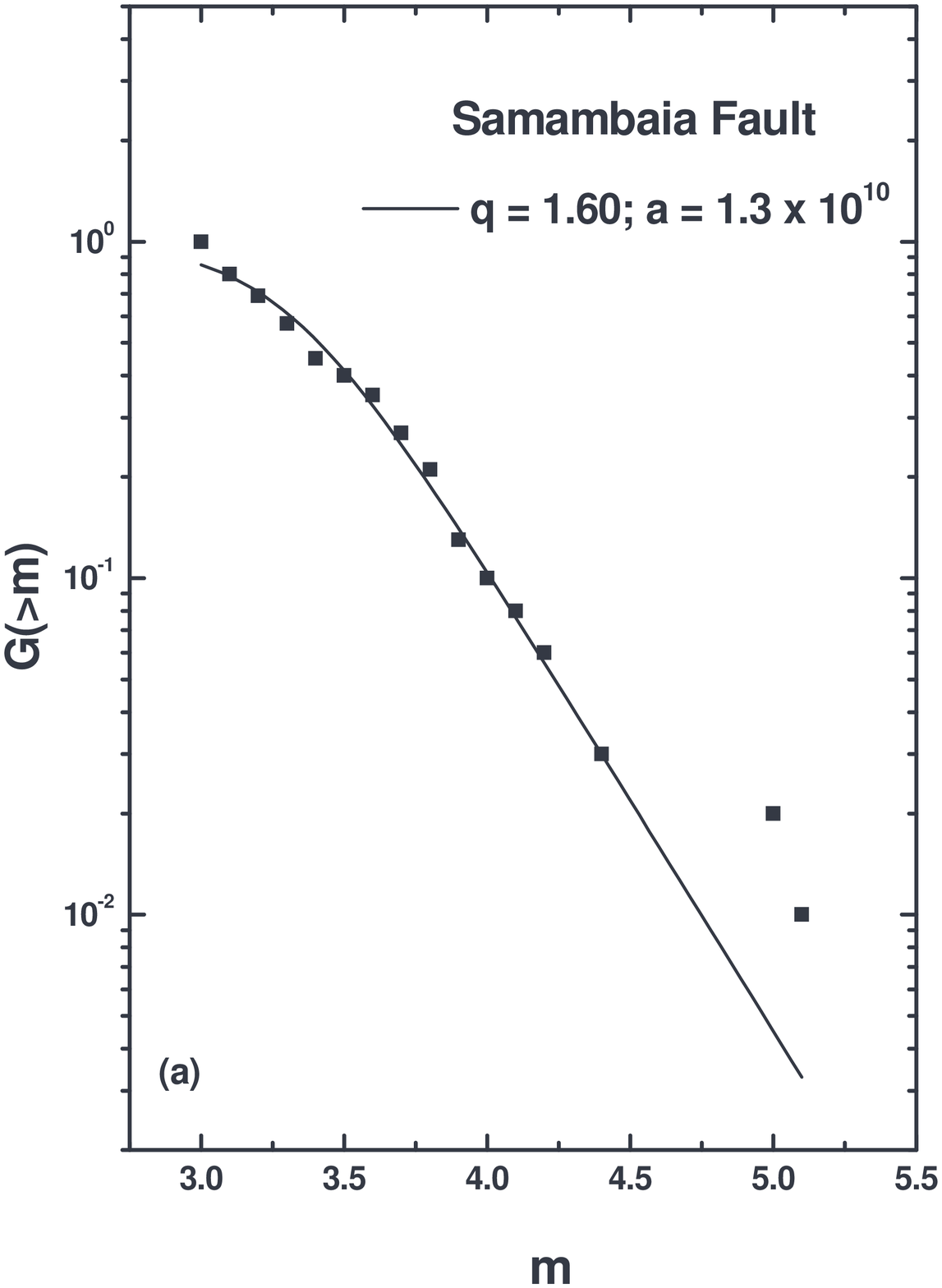,width=2.4truein,height=4.1truein}
\psfig{figure=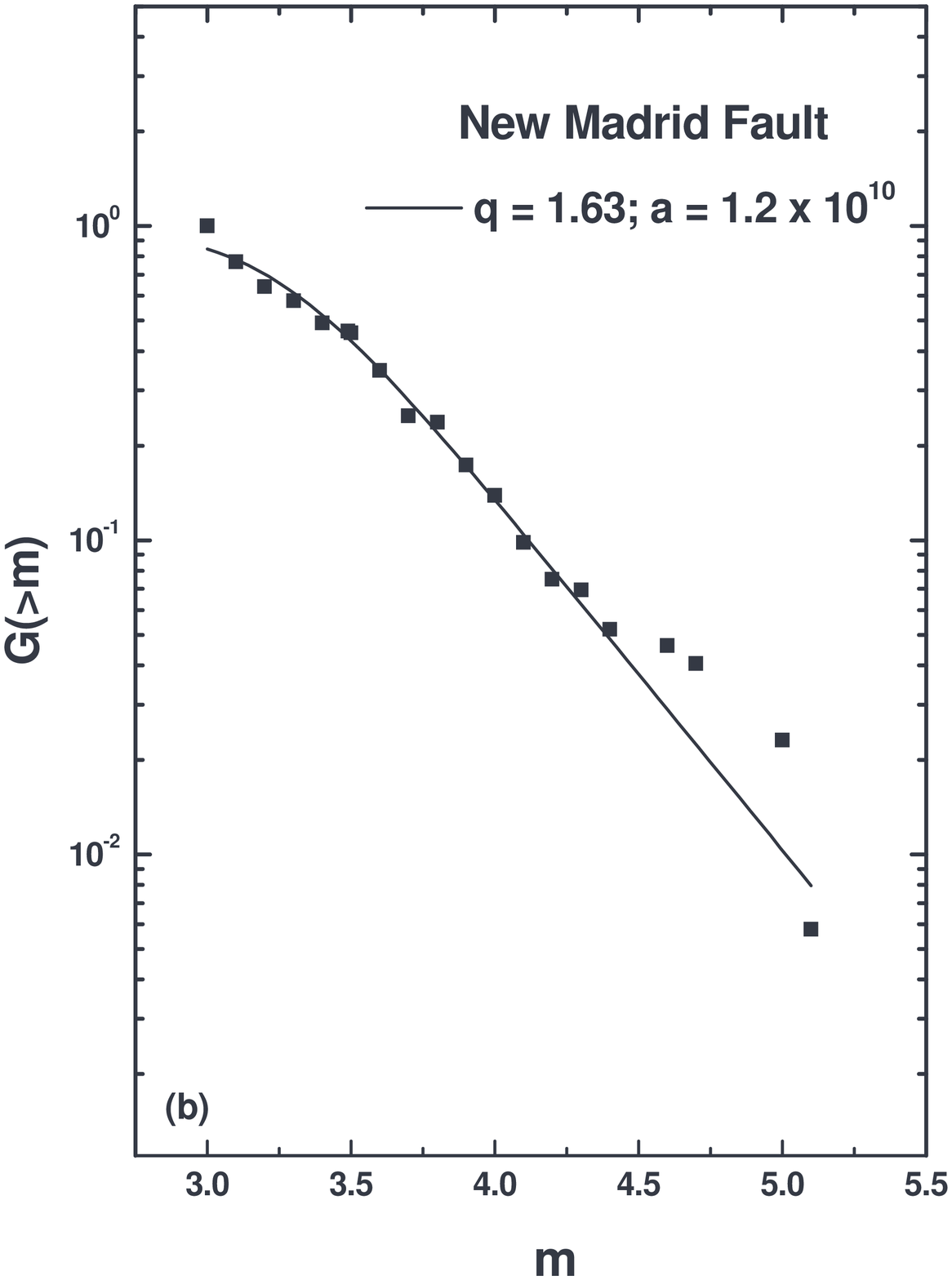,width=2.4truein,height=4.1truein}
\psfig{figure=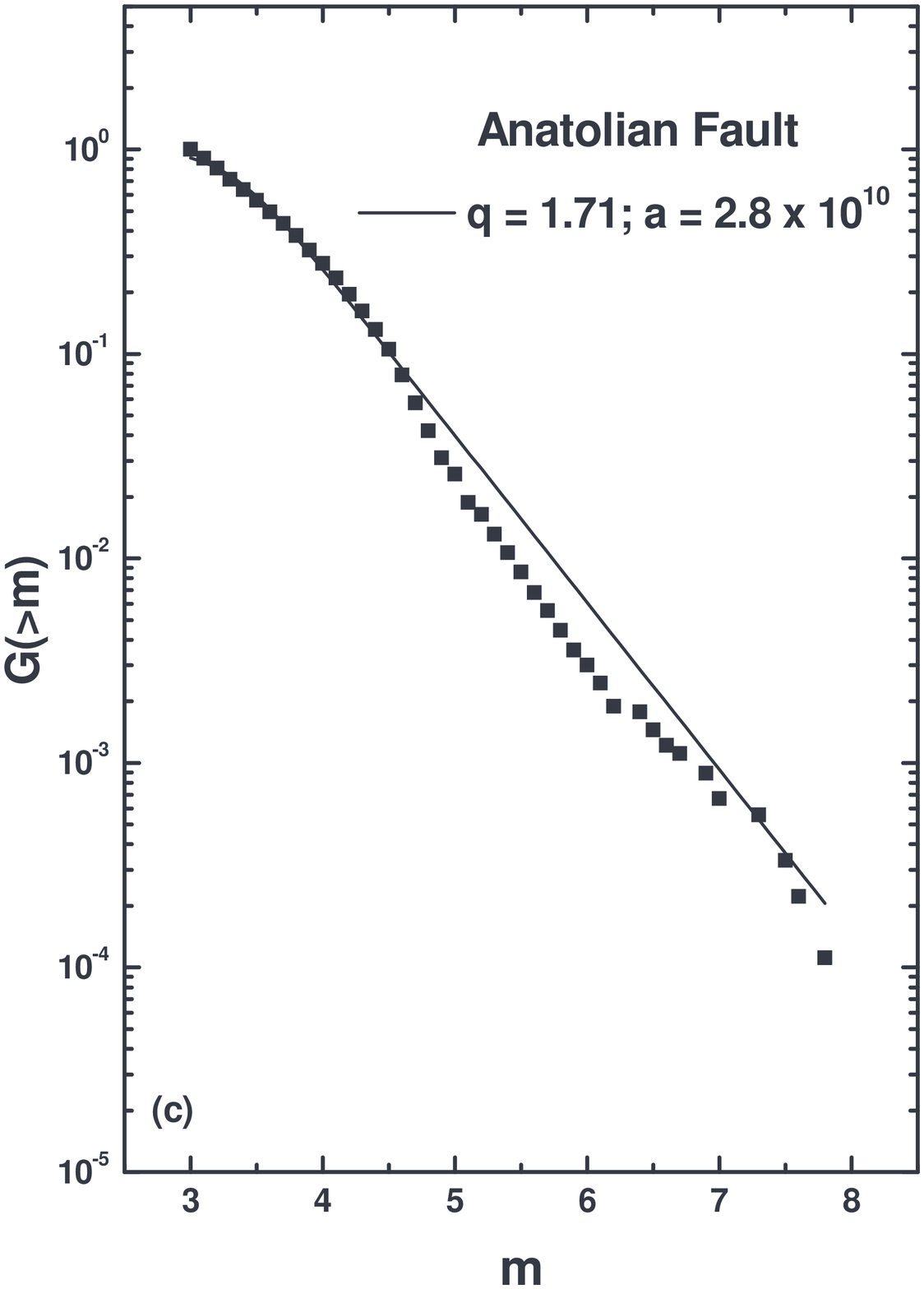,width=2.4truein,height=4.1truein}
\hskip 0.1in}
\caption{The relative cumulative number of earthquakes [Eq. (\ref{11})] as a function of the magnitude $m$. In all panels, 
the data points correspond to earthquakes lying in the interval $3 < m < 8$. {\bf{(a):}} 
Samambaia fault - Brazil: 100 events from Bulletin Seismic of the Revista Brasileira de Geof\'{\i}sica. 
{\bf{(b):}} New Madrid fault - USA: 173 data points taken from NEIC catalog. {\bf{(c):}} 
Anatolian fault - Turkey: 8980 events from NEIC catalog. The best-fit values for the 
parameters $q$ and $a$ are shown in the respective Panel. A summary of this analysis is also shown in Table I.}
\end{figure*}

\section{New approach}

Now, let us discuss the standard method of maximization of the Tsallis entropy. Here and hereafter, the Boltzmann 
constant is set equal to unity for the sake of simplicity. Thus, the functional entropy to be maximized is
\begin{equation}
\delta S^*_q =\delta\left(S_q +\alpha\int_0^\infty p(\sigma)d\sigma - \beta \sigma_q\right) = 0,
\end{equation}
where $\alpha$ and $\beta$ are the Lagrange multipliers. The constrains used above are the normalization of the 
distribution
\begin{equation}
\int_0^{\infty} p(\sigma) d\sigma = 1
\end{equation}
and {\it the $q$-expectation value}
\begin{equation}
\sigma_q=<\sigma>_q=\int_0^{\infty} \sigma P_q (\sigma) d\sigma
\end{equation}
with the escort distribution \cite{abe03} given by
\begin{equation}
P_q={p^q(\sigma) \over \int_0^{\infty} p^q(\sigma) d\sigma}.
\end{equation}
By considering the same physical arguments of Ref. \cite{oscar2004}, we derive, after some algebra,
the following expresion for the fragment size distribution function
\begin{equation}\label{sigma}
p(\sigma)= \left[1-{(1-q)\over (2-q)}(\sigma - \sigma_q)\right]^{1 \over 1-q},
\end{equation}
which corresponds to the area distribution for the fragments of the fault plates. Here, however, differently 
from Ref. \cite{oscar2004}, which assumes $\varepsilon \sim r$, we use a new energy scale $\varepsilon \sim r^3$. 
Thus, the proportionality between the released relative energy $\epsilon$ and $r^3$ ($r$ is the size of fragments) 
is now given by $\sigma - \sigma_q = {\left(\varepsilon/a\right)^{2/3}}$, where $\sigma$ scales with $r^2$ and $a$ (the 
proportionality constant between $\varepsilon$ and $r^3$) has dimension of volumetric energy density. 
In particular, this new scale is in accordance with the standard theory of seismic rupture, 
the well-known seismic moment scaling with rupture length (see, for instance \cite{thorne95}).

The new EDF of earthquakes is, therefore, obtained by changing variables from $\sigma-\sigma_q$ to 
${\left(\varepsilon/a\right)^{2/3}}$. From (\ref{sigma}) it is straightforward to show that
\begin{equation}\label{pe}
p(\varepsilon )d\varepsilon = \frac{C\varepsilon^{-{1\over 3}} d\varepsilon}{\left[1 + C'\varepsilon^{2/3}\right]^{1 \over q-1}},
\end{equation}
which has also a power-law form with $C$ and $C'$ given by
\begin{equation}
C={2\over 3 a^{2/3}}\quad{\rm and}\quad C'=-{(1-q)\over(2-q)a^{2/3}}.
\end{equation}
In the above expression, the energy probability is written as $p(\varepsilon)=n(\varepsilon)/N$, where $n(\varepsilon)$ 
corresponds to the number of earthquakes with energy $\varepsilon$ and $N$ total number earthquakes.

\section{Testing the new EDF with the cumulative number of earthquakes}

In order to test the viability of the new EDF above derived [Eq. (\ref{pe})] we introduce the cumulative number of 
earthquakes, given by integral \cite{oscar2004}
\begin{equation}\label{inte}
{N_{\epsilon >}\over N} =\int _{\varepsilon}^\infty p(\varepsilon) d\varepsilon,
\end{equation}
where $N_{\varepsilon >}$ is the number of earthquakes with energy larger than $\varepsilon $. Now, substituting
(\ref{pe}) into (\ref{inte}), and considering $m = \frac{1}{3} \log\varepsilon$ ($m$ stands for magnitude) 
it is possible to calculate the above expression. In reality, note that depending on the value of $q$ 
the limits of the integral (\ref{inte}) presents a cutoff on the maximum value allowed for energy $\epsilon$, 
which is given by $\epsilon_{max}=\sqrt{a^{2/3}(2-q)/(1-q)}$ for the intervals $q<1$ and $q>2$, while 
for $1 < q < 2$ the cutoff is absent in the distribution. Note also that in the limit $q\rightarrow 1$, 
$\epsilon_{max} \rightarrow \infty$ and $p(\varepsilon)$ goes to the exponential function. As matter of fact, 
the calculation of the integral (\ref{inte}) for $q\neq 1$ leads to the general expression 
\begin{eqnarray}\label{11}
\log (N_{>m}) & = & \log N + \left(\frac{2-q}{1-q}\right) \times \\ \nonumber & & \times \log \left[1 -
\left(\frac{1-q}{2-q}\right)\times \left({10^{2m}\over a^{2/3}}\right)\right],  
\end{eqnarray}
which, similarly to the modified Gutenberg-Ricther law (See, e.g., Refs. \cite{sornette} for more details), describes appropriately the energy distribution in a wider detectable range of magnitudes.

Figure 1 shows the relative cumulative number of earthquakes ($G_{m>}=N_{m>}/N$) as a function of the magnitude 
$m$. The data points, corresponding to earthquakes events lying in the 
interval $3 < m < 8$, were taken from two different catalogs, namely, Bulletin Seismic of the Revista Brasileira de 
Geof\'{\i}sica (left panels) and NEIC (central and right panels). The left, central and right Panels show 
the results of our analysis for the Samambaia fault, Brazil (100 events), New madrid fault, USA (173 events), and 
Anatolian fault, Turkey (8980 events), respectively.  We note that, similarly to original version of SCP model, 
our approach, represented by Eqs. (\ref{sigma})-(\ref{11}), provide a very good fit to the experimental data of the 
two catalogs here considered. It is worth emphasizing, however, that the energy density differ by several orders of 
magnitude from our model to the original SCP model. Therefore, we expect that other independent 
estimates of the parameter $a$ may indicate which approach is more physically realistic. The estimates of the 
parameters $q$ and $a$ obtained in this paper and in Ref. \cite{oscar2004} are summarized in Table 1.

\begin{table}[t]
\caption{Limits to $q$ and $a$}
\begin{ruledtabular}
\begin{tabular}{lclc}
Fault& Ref. &\quad \quad $q$ & \quad $a$ \\
\hline \hline \\
California - USA & \cite{oscar2004} & \quad $1.65$& \quad $5.73 \times 10^{-6}$\\
Iberian Penisula - Spain & \cite{oscar2004} & \quad $1.64$& \quad $3.37 \times 10^{-6}$\\
Andaluc\'{\i}a - Spain & \cite{oscar2004} & \quad $1.60$& \quad $3.0 \times 10^{-6}$\\
\hline \\
Samambaia - Brazil & This Paper & \quad $1.60$& \quad $1.3 \times 10^{10}$\\
New Madrid - USA & This Paper & \quad $1.63$ & \quad $1.2 \times 10^{10}$\\
Anatolian - Turkey & This Paper & \quad $1.71$ & \quad $2.8 \times 10^{10}$\\
\end{tabular}
\end{ruledtabular}
\end{table}

\section{conclusion}

In Ref. \cite{abe05}, what seems to be the correct definition for expectation values within the Tsallis 
nonextensive statistical mechanics was rediscussed. Based on properties of the generalized relative entropies and 
the Shore-Johnson theorem, it was shown that the expectation value of any physical quantity in this extended framework 
converges to the normalized $q$-expectation value, instead of to the ordinary definition.

In this paper, by considering this \emph{necessity} of $q$-expectation values in Tsallis nonextensive framework, 
we have revisited the fragment-asperity interaction model for earthquakes, as introduced in Ref. \cite{oscar2004}. 
A new energy distribution function has been calculated, which allowed us to determine the relative cumulative number 
of earthquakes as a function of the magnitude. Additionally, a new scale law between the released relative 
energy $\epsilon$ and the volume of fragments $r^3$ has also been introduced, i.e., in agreement with the so-called seismic moment scaling 
with rupture length. As discussed earlier, although our analysis and 
the one presented in Ref. \cite{oscar2004} provide very similar values for the nonextensive parameter $q$, the other 
physical quantity, e.g., the energy density differ by several orders of magnitude. It would be interesting, 
therefore, if we could have experimental estimates for these quantities in order to compare the predictions of the models. 
Finally, it is worth mentioning that the estimates for the nonextensive parameter from the two catalogs here considered 
(Fig. 1) are consistent with the upper limit $q<2$, obtained from several independent studies involving the Tsallis 
nonextensive framework \cite{newref}.

{\it Acknowledgments:} The authors thank the anonymous referees for their valuable suggestions and comments. We also thank the partial support by the Conselho Nacional de Desenvolvimento Cient\'{\i}fico e Tecnol\'ogico (CNPq - Brazil). JSA is supported by CNPq (307860/2004-3) and CNPq (475835/2004-2). CSV is supported by FAPERN.

\end{document}